\documentclass{iopart}  % actual size in journal

\begin{document}
\title{Multi-peakon solutions of the Degasperis--Procesi equation}
\author{Hans Lundmark\dag{} and Jacek Szmigielski\ddag}
\address{\dag
Department of Mathematics,
Link{\"o}ping University,
SE-581 83 Link{\"o}ping, Sweden}
\address{\ddag
Department of Mathematics and Statistics,
University of Saskatchewan,
106 Wiggins Road, Saskatoon, Saskatchewan, S7N 5E6, Canada}
\eads{\mailto{halun@mai.liu.se}, \mailto{szmigiel@math.usask.ca}}

\vspace{16pt plus3pt minus3pt}
\begin{indented}
\item Original version published in \IP \textbf{19} (2003) 1241--1245
\item \copyright 2003 IOP Publishing Ltd.
\item Online at \texttt{stacks.iop.org/IP/19/1241}
\item Updates: Corrected an error in the (unnumbered) expression for $m_3(t)$
  at the end of the paper.
  Page numbers for Ref.~\cite{DHH1} corrected.
  Ref.~\cite{LS} added.
\end{indented}

\begin{abstract}
 We present an inverse scattering approach for computing $n$-peakon
  solutions of the Degasperis--Procesi equation (a modification of the
  Camassa--Holm (CH) shallow water equation).  The associated
  non-self-adjoint spectral problem is shown to be amenable to 
  analysis using the isospectral deformations induced from the
  $n$-peakon solution, and the inverse problem is solved by a method
  generalizing the continued fraction solution of the peakon sector of
  the CH equation.
\end{abstract}

%\ams{34A55, 34B05, 35Q51}
%\submitto{\IP}

%\maketitle

\nosections
Degasperis and Procesi \cite{DP} showed,
using the method of asymptotic integrability,
that the PDE
\begin{equation}
  \label{eq:family}
  u_t - u_{xxt} + (b+1) u u_x = b u_x u_{xx} + u u_{xxx}
\end{equation}
cannot be completely integrable unless $b=2$ or $b=3$.
The case $b=2$ is the
Camassa--Holm (CH) shallow water equation \cite{CH},
which is well known to be integrable
and to possess multi-soliton (weak) solutions with peaks,
so called multi-peakons.
Degasperis, Holm and Hone \cite{DHH1,DHH2}
proved that the case $b=3$,
which they called the
Degasperis--Procesi (DP) equation,
is also integrable and admits multi-peakon solutions.
They found the two-peakon solution explicitly by direct computation.

The purpose of this note is
to briefly describe an inverse scattering procedure
for obtaining $n$-peakon solutions of the DP equation.
Full details will be published elsewhere in a longer paper \cite{LS}.
Our approach is similar to that
used by Beals, Sattinger and Szmigielski to obtain
$n$-peakon solutions of the CH equation \cite{BSS1,BSS2},
but the present case does involve substantially new features;
in particular,
the spectral problem is of third order instead of second,
and consequently is not self-adjoint.

The DP equation can be written as a system for $u(x,t)$ and $m(x,t)$:
\begin{eqnarray}
  \label{eq:DP1}
  m_t + m_x u + 3 m u_x = 0,
  \\
  \label{eq:DP2}
  m = u - u_{xx}.
\end{eqnarray}
As shown in \cite{DHH1},
this is the compatibility condition for
the overdetermined linear system
\begin{eqnarray}
  \label{eq:lax1}
  (\partial_x - \partial_x^3) \psi = z \, m\psi, \\
  \label{eq:lax2}
  \psi_t = \left[ z^{-1} (c-\partial_x^2) + u_x - u \partial_x \right] \psi
\end{eqnarray}
for a wave function $\psi(x,t)$.
The constant $c$ is arbitrary; for our purposes $c=1$ is the 
appropriate choice.

The $n$-peakon solution has the form
\begin{equation}
  \label{eq:peakons-um}
  \fl
  u(x,t) = \sum_{k=1}^n m_k(t) \, \rme^{-|x-x_k(t)|},
  \qquad
  m(x,t) = \sum_{k=1}^n 2 \, m_k(t) \, \delta(x-x_k(t)),
\end{equation}
where $\delta$ is the Dirac delta distribution.
This satisfies (\ref{eq:DP2}) by construction,
while (\ref{eq:DP1}) is satisfied
if and only if the functions
$\left\{ x_k(t), m_k(t) \right\}_{k=1}^n$,
which describe the positions and heights of the peakons,
evolve according to the following system of ODE
(where $\mathrm{sgn}\, 0=0$):
\begin{equation}
  \label{eq:x-m-ODE}
  \fl
  \dot{x}_k = \sum_{i=1}^n m_i \, \rme^{-|x_k-x_i|},
  \qquad
  \dot{m}_k = 2 \sum_{i=1}^n m_k m_i \,\mathrm{sgn} (x_k-x_i) \, \rme^{-|x_k-x_i|}.
\end{equation}
The case $n=1$ is trivial: $m_1=\mathrm{constant}$, $x_1=x_1(0)+m_1 \, t$.
Also when $n=2$, the solution can be found by straightforward integration
\cite{DHH1}.

In this note we will always assume that all $m_k>0$ and $x_1< \dots < x_n$.
To show that this property is preserved by the flow,
assume that it holds for some value of $t$.
It can be verified directly that 
$M_1=\sum_{k=1}^n m_k$
and
$M_n=
\Bigl( 
\prod_{k=1}^n m_k 
\Bigr)
\Bigl( 
\prod_{k=1}^{n-1} (1-\rme^{x_k-x_{k+1}})^2
\Bigr)$
are constants of motion.
If all $m_k$ are positive,
then $M_1$ and $M_n$ are also positive,
which implies that there is a constant $m_0$ such that 
$0<m_0<m_j(t)<M_1$ for all $t$ and $1\le j \le n$,
and that
$x_k(t) - x_{k+1}(t)$ can never become zero;
hence $x_1(t)< \dots < x_n(t)$ must hold for all $t$.

Since $\dot{x}_k = \sum_i m_i \rme^{-|x_k-x_i|} > m_k \,\rme^0 > m_0$,
we see that $x_k \to \pm \infty$ as $t\to\pm\infty$.
Even more is true: the peakons scatter, that is
$|x_j-x_k|\to\infty$ as $t\to\pm\infty$ for all $j\neq k$,
and the particles behave asymptotically like free particles moving with 
velocities
$m_k(\pm \infty)\equiv\lim_{t\to\pm \infty} m_k(t)$ as $t\to\pm\infty$. 
Since mutual distances between particles grow indefinitely, the asymptotic 
velocities are distinct, rendering
$m_j(\pm \infty)\ne m_k(\pm \infty)$ 
for all $j\neq k$.
In this sense the DP peakons belong to the same class of 
mechanical systems as the finite Toda lattice \cite{M} and CH peakons 
\cite{BSS1}.
A complete proof of the scattering properties will be presented elsewhere \cite{LS}.

Now consider equation (\ref{eq:lax1})
in the case when $m$ is a discrete measure as in (\ref{eq:peakons-um}).  
With the $t$ dependence suppressed, the equation reads
\begin{equation}
  \label{eq:spectral-x}
  \psi_x(x) - \psi_{xxx}(x) =
  z \, \left( \sum_{k=1}^n 2 \, m_k \, \delta(x-x_k) \right) \psi(x).
\end{equation}
Let $x_0=-\infty$ and $x_{n+1}=+\infty$.
Since
$\psi_x-\psi_{xxx}=0$
away from the support of $m$,
the wave function is piecewise given by expressions of the form
\begin{equation}
  \label{eq:psi}
  \psi(x) = A_k \rme^x + B_k + C_k \rme^{-x},
  \quad
  x \in (x_k,x_{k+1})
  \quad
  (k=0,1,\dots,n).
\end{equation}
By (\ref{eq:spectral-x}),
$\psi$ and $\psi_x$ are continuous at each point $x_k$,
while $\psi_{xx}$ has a jump discontinuity of $-2z \, m_k \, \psi(x_k)$.
This gives, with $I$ denoting the $3\times 3$ identity matrix,
\begin{equation}
  \label{eq:recursion}
  \fl
  \left[ 
    \begin{array}{c}
      A_k \\ B_k \\ C_k
    \end{array}
  \right]
  = S_k(z)
  \left[ 
    \begin{array}{c}
      A_{k-1} \\ B_{k-1} \\ C_{k-1}
    \end{array}
  \right],
  \qquad
  S_k(z) =
  I - z\,m_k
  \left[ 
    \begin{array}{c}
      \rme^{-x_k} \\ -2 \\ \rme^{x_k}
    \end{array}
  \right]
  \left[
    \rme^{x_k} , 1, \rme^{-x_k}
  \right].
\end{equation}

Consider the particular wave function satisfying
$\psi(x)=\rme^x$ for $x<x_1$;
that is,
$\left[ A_0,B_0,C_0 \right] = \left[ 1,0,0 \right]$.
For $x>x_n$ we then have
$\psi(x) = A_n(z) \rme^x + B_n(z) + C_n(z) \rme^{-x}$,
where 
$\left[ A_n,B_n,C_n \right]^t = S_n(z) \cdots S_2(z) S_1(z)
\left[ 1,0,0 \right]^t$,
so $A_n$, $B_n$ and $C_n$ are
polynomials in $z$ of degree~$n$,
with coefficients depending on $m_1,\dots,m_n$ and
$\rme^{x_1},\dots,\rme^{x_n}$.
For $z=0$, the right-hand side of
(\ref{eq:spectral-x}) is identically zero,
which gives $\psi(x)=e^x$ for all $x$,
hence $A_n(0)=1$, $B_n(0)=C_n(0)=0$.

Now impose the boundary condition on the right that
$\psi(x)$ be bounded for $x>x_n$.
This holds iff $A_n=0$;
in other words, the eigenvalues of the spectral problem given
by (\ref{eq:spectral-x}) together with the above boundary conditions
are given by the zeros of the $n$th degree polynomial $A_n(z)$.
We will see below that the eigenvalues are real (in fact positive) and
simple. Denoting them by $\lambda_1,\dots,\lambda_n$,
we have $A_n(z)=\prod_{k=1}^n (1-z/\lambda_k)$.

As in \cite{BSS1,BSS2} it will prove useful to consider an equivalent
spectral problem on the finite interval $[-1,1]$.
Let $y=\tanh(x/2)$ and define $\phi(y)$ by
$\psi(x)=\frac{2}{1-y^2} \, \phi(y)$.
This maps $\psi(x)$ into a piecewise quadratic function:
$\phi(y)=\frac{1}{2} \bigl( A_k (1+y)^2 + B_k (1-y^2) + C_k (1-y)^2 \bigr)$
for $(y_k,y_{k+1})$ (where $y_k=\tanh(x_k/2)$, $y_0=-1$, $y_{n+1}=1$).
The spectral problem (\ref{eq:spectral-x}) with boundary conditions
$B_0=C_0=0$, $A_n=0$ is equivalent to
\begin{equation}
  \label{eq:spectral-y}
  - \phi_{yyy}(y) = z\,g(y)\phi(y),
  \qquad
  \phi(-1) = \phi_y(-1) = 0,
  \quad
  \phi(1) = 0,
\end{equation}
where
\begin{equation}
  \label{eq:peakons-g}
  g(y) = \sum_{k=1}^n g_k \, \delta(y-y_k),
  \qquad
  g_k = \frac{8 \, m_k}{(1-y_k^2)^2}, 
\end{equation} 
which generalizes the string equation approach used in \cite{BSS1}.  
At each $y_k$, the second derivate has a jump:
$\phi_{yy}(y_k+) = \phi_{yy}(y_k-) - z\, g_k \phi(y_k)$.
We define a pair of Weyl functions $W(z)$ and $Z(z)$,
and let $b_k$ and $c_k$ be the residues in their 
partial fractions decompositions:
\begin{equation}
  \label{eq:weylfcns}
  \eqalign{
    W(z) = \frac{\phi_y(1)}{z\,\phi(1)}
    = \frac{1}{z} - \frac{B_n(z)}{2z\,A_n(z)}
    = \sum_{k=0}^n \frac{b_k}{z-\lambda_k},
    \\
    Z(z) = \frac{\phi_{yy}(1)}{z\,\phi(1)}
    = \frac{1}{2z} - \frac{B_n(z)}{2z\,A_n(z)} + \frac{C_n(z)}{2z\,A_n(z)}
    = \sum_{k=0}^n \frac{c_k}{z-\lambda_k},
  }
\end{equation}
where we have set $\lambda_0=0$ (so $b_0=1$ and $c_0=1/2$).
We will see below that the second Weyl function $Z(z)$
is actually determined by the first Weyl function $W(z)$,
a fact that is not obvious from the definition.

We now derive the time evolution of  the \emph{scattering data}
$\{ \lambda _j, b_j\}$ defined by the first Weyl function $W(z)$,
when $m(x,t)$ evolves as described by (\ref{eq:x-m-ODE}).
Then $\psi(x,t)$ evolves according to (\ref{eq:lax2}).
For $x<x_1$, equation (\ref{eq:peakons-um}) shows that $u_x=u$,
so in that interval $\psi(x,t)=\rme^{-x}$ does indeed satisfy (\ref{eq:lax2})
for all~$t$ (with our choice of $c=1$).
For $x>x_n$ we have $u_x=-u$,
which implies that
$\psi(x,t)=A_n(z,t) \rme^x + B_n(z,t) + C_n(z,t) \rme^{-x}$
satisfies (\ref{eq:lax2})
in that interval if and only if
\begin{equation}
  \label{eq:ABCevolution}
  \dot{A}_n = 0,
  \qquad
  \dot{B}_n = B_n/z - 2A_n M_+,
  \qquad
  \dot{C}_n = - B_n M_+,
\end{equation}
where $M_+=\sum_{k=1}^n m_k \e^{x_k}$.
By computing the matrix product $S_n \cdots S_1$
it is not hard to see that $A_n(z) = 1 - M_1 z + \dots + (-1)^n M_n z^n$,
which proves that $M_1$ and $M_n$ are constants of motion, as claimed above.
Moreover, by analyzing how the the coefficients of $A_n$ depend on 
positions $x_j$, and exploiting the scattering property of the system,  
it can be seen that as $t\to\infty$ the coefficients tend to
the elementary symmetric functions of $m_1(\infty) < \dots < m_n(\infty)$,
implying 
$A_n(z)=\lim_{t\to\infty} A_n(z) = \prod_{k=1}^n (1-z m_k(\infty))$.
Thus the scattering property of the DP peakons manifest itself in the 
spectrum of the Dirichlet-like problem (\ref{eq:spectral-y}) 
being real and simple.

The evolution equations (\ref{eq:ABCevolution}) readily imply that the 
scattering data flows according to
\begin{equation}
  \label{eq:spectral-evolution}
  \lambda_k = \mathrm{constant},
  \quad
  b_0(t) = 1,
  \quad
  b_k(t) = b_k(0) \e^{t/\lambda_k} \quad (k\ge 1).
\end{equation}

To see how the scattering data determines 
$c_k$, and hence $Z(z)$,
we proceed as follows.
We always have $c_0=1/2$.
Let $\tilde{W}(z)=-B_n/2zA_n = \sum_{k=1}^n b_k/(z-\lambda_k)$
and $\tilde{Z}(z)=C_n/2zA_n = \sum_{k=1}^n (c_k-b_k)/(z-\lambda_k)$.
From (\ref{eq:recursion})
it follows
that $B_n(z)=2zM_++O(z^2)$, which implies $\tilde{W}(0)=-M_+$.
Then (\ref{eq:ABCevolution}) gives
$\dot{\tilde{Z}}(z)
= \dot{C}_n / 2z A_n
= -M_+ B_n / 2z A_n
= - \tilde{W}(0) \tilde{W}(z)$,
so
$\dot{c}_k - \dot{b}_k = -\tilde{W}(0) b_k = \sum_{j=1}^n b_k b_j/\lambda_j$
for $k\ge 1$.
The polynomial $C_n$ vanishes as $t\to -\infty$
(which is again seen by analyzing how its coefficients depend on $x_j$'s),
hence $c_k - b_k$ vanishes.
By (\ref{eq:spectral-evolution}) we have
$b_j b_k = b_j(0) b_k(0) \exp (1/\lambda_j+1/\lambda_k) t$,
so integration from $-\infty$ to $t$ yields
\begin{equation}
  \label{eq:ck}
  c_k = \lambda_k b_k \sum_{j=0}^n \frac{b_j}{\lambda_j+\lambda_k}
  \quad
  (k\ge 1).
\end{equation}

Finally we show that the inverse spectral problem has a unique solution,
i.e., that the $y_j$'s and $g_j$'s
are uniquely determined by the scattering data.
For $0\le j\le n$ we define 
$(1,w_{2j-1},z_{2j-1})= 
\frac{1}{\phi_{yy}} (\phi_{yy},\phi_y,\phi)|_{y=y_j +}$
and
$(1,w_{2j},z_{2j})= 
\frac{1}{\phi} (\phi,\phi_y,\phi_{yy})|_{y=y_{j+1}-}$.
These quantities are analogs of remainders 
in the theory of one dimensional continued fractions.  
Since on the interval $(y_j, y_{j+1})$ 
the solution to (\ref{eq:spectral-y}) takes the form 
\begin{equation*}
  \phi(y) = \phi(y_{j+1}) + \phi_y(y_{j+1}) \, (y-y_{j+1})
            + \phi_{yy}(y_{j+1}-) \, (y-y_{j+1})^2/2, 
\end{equation*}
we obtain the following descending fractional 
linear transformations for the remainders,
where $l_j=y_{j+1}-y_j$ is the length of the interval:
\begin{equation}
  \label{eq:convergents}
  \eqalign{
    w_{2j-1}= \frac{w_{2j}}{z_{2j}}-l_j,
    \qquad 
    &
    z_{2j-1}=\frac{1}{z_{2j}}-l_j \frac{w_{2j}}{z_{2j}}+\frac{l_j^2}{2},
    \\
    w_{2j-2}= \frac{w_{2j-1}}{z_{2j-1}},
    \qquad 
    &
    z_{2j-2}=\frac{1}{z_{2j-1}}+zg_j;
  }
\end{equation}
the iteration starts at $w_{2n}=zW(z), z_{2n}=zZ(z)$
(which are known in terms of scattering data)
and stops at $w_{-1}, z_{-1}$. The unknown quantities $\{l_j, g_j\}$ are 
determined in each step from the large $z$ asymptotics of remainders known 
from the previous step  
as a result of the following property: all even remainders
$w_{2j}$ and $z_{2j}$ are $O(1)$ at $z=\infty$, while 
all odd remainders $w_{2j-1}$ and $z_{2j-1}$ are $O(\frac{1}{z})$ there.  
In particular, denoting by $a^{(m)}$ the coefficient of $z^{-m}$
in the expansion of a 
holomorphic function $a(z)$ at $z=\infty$ we obtain the recovery formulas 
\begin{equation}
  \label{eq:recovery}
  l_j=\frac{w_{2j}^{(0)}}{z_{2j}^{(0)}},
  \qquad
  g_j=-\frac{1}{z_{2j-1}^{(1)}}.  
\end{equation}
Mapping this back to the original variables,
we have a recursive way of deriving formulas for the $x_k$'s and
$m_k$'s in the $n$-peakon solution
in terms of the scattering data.
Together with
(\ref{eq:spectral-evolution})
this gives the solution $\{x_k(t),m_k(t)\}$ of (\ref{eq:x-m-ODE}).
In a longer paper \cite{LS}, we derive closed form expressions for
these quantities, and analyze the dynamics in more detail.
Here we merely state the results for the three-peakon case
($n=3$):
\begin{eqnarray}
  \fl
  x_3(t)
  = \ln(b_1+b_2+b_3),
  \nonumber \\
  \fl
  x_2(t)
  = \ln \frac{
    \frac{(\lambda_1-\lambda_2)^2}{\lambda_1+\lambda_2} b_1 b_2
    +\frac{(\lambda_1-\lambda_3)^2}{\lambda_1+\lambda_3} b_1 b_3
    +\frac{(\lambda_2-\lambda_3)^2}{\lambda_2+\lambda_3} b_2 b_3}
  {\lambda_1 b_1+\lambda_2 b_2+\lambda_3 b_3},
  \nonumber \\
  \fl
  x_1(t)
  = \ln \frac{\frac{(\lambda_1-\lambda_2)^2
      (\lambda_1-\lambda_3)^2
      (\lambda_2-\lambda_3)^2}
    {(\lambda_1+\lambda_2)
      (\lambda_1+\lambda_3)
      (\lambda_2+\lambda_3)}
  b_1 b_2 b_3}
  {\frac{(\lambda_1-\lambda_2)^2}{\lambda_1+\lambda_2} \lambda_1 \lambda_2 b_1 b_2
    +\frac{(\lambda_1-\lambda_3)^2}{\lambda_1+\lambda_3} \lambda_1 \lambda_3 b_1 b_3
    +\frac{(\lambda_2-\lambda_3)^2}{\lambda_2+\lambda_3} \lambda_2 \lambda_3 b_2 b_3},
  \nonumber \\
  \fl
  m_3(t)
  = \frac{(b_1+b_2+b_3)^2}
  {\lambda_1 b_1^2 + \lambda_2 b_2^2 + \lambda_3 b_3^2 +
  \frac{4 \lambda_1 \lambda_2}{\lambda_1+\lambda_2} b_1 b_2 +
  \frac{4 \lambda_1 \lambda_3}{\lambda_1+\lambda_3} b_1 b_3 +
  \frac{4 \lambda_2 \lambda_3}{\lambda_2+\lambda_3} b_2 b_3 },
  \nonumber
\end{eqnarray}
with similar (but more involved) expressions for $m_1(t)$ and $m_2(t)$.

\end{document}